# Educating Reflective Systems Developers at Scale: Towards "productive feedback" in a semi-capstone large-scale software engineering course[1]

Torgeir Dingsøyr
Department of Computer Science
*Norwegian University of Science and Technology*
Trondheim, Norway
0000-0003-0725-345X

*Abstract*—Feedback is critical in education. This Innovative Practice Full Paper reports lessons learned from improving the quality of feedback in a semi-capstone software engineering course, with particular focus on how to deliver productive feedback in large scale during project work. The bachelor-level introduction to software engineering course is taken by about 500 students from eight study programs, organised into 72 project teams. The course aims to educate reflective systems developers. The teaching staff includes 29 teaching assistants as supervisor and product owners for teams. Project teams get feedback on seven deliverables as part of formative portfolio assessment. Students expressed frustration on feedback not being aligned, that they got critique on topics not stated in assignments and that teaching assistants were reluctant to discuss the feedback. This article provides a description of the course design, an assessment of the quality of feedback and lessons learned from three main changes: Revising assignments and rubrics, reorganising the teaching staff and increasing training of teaching assistants. In discussing the changes, we draw on a survey to students with 142 respondents, a survey to teaching assistants with 18 respondents, meeting minutes from a student reference group and experience reports from teaching assistants as well as literature and own experience. The article concludes with three actionable lessons learned for large-scale semi-capstone courses.

## I. Introduction

In the autumn of 2020 I started in a new position where I was asked to be course responsible for a large bachelor-level course in software engineering which was given in the spring of 2021. The course aims to provide students with basic knowledge in software engineering, in that they after having completed the course should be able to plan and manage small development projects and participate in activities such as programming, testing, and leading project work. Students should also develop understanding of software development as a profession, and be able to reflect on complex software development projects with technical and organisational challenges.

The course is given in the fourth semester to eight study programs and has 7.5 ECTS. It has been taught in several ways the last ten years [1, 2], as traditional lectures supported by a textbook, to a flipped classroom approach, to the current set-up where project work is a central part. In the teaching staff, we share the ambition which Mathiassen and Puaro describe as to educate *reflective systems developers* who "*must ... bring to bear something more than a repertoire of general methods and tools. They must engage in reflections and dialogues to generate the necessary insights into the situation at hand*" [3].

In 2021, the course had about 500 students, organised into 72 teams. The teaching staff included 29 teaching assistants and two lecturers. The course was divided into four parts:

i) Theory-module: Interactive online sessions and individual reading of curriculum (2 weeks).

ii) Project work, first iteration (5 weeks).

iii) Project work, second iteration (5 weeks).

iv) Reflection in teams (3 weeks).

---

The theory module provided an overview of software engineering with main focus on development process and Scrum, but also covered requirements engineering, software architecture and software quality. Students read articles, parts of two textbooks [4, 5] and could participate in online training sessions combining brief lectures with practical exercises and small discussions. The theory module ended with a quiz, but which did not impact on the final grade.

In the second and third part, student teams with on average 7 members were assigned with the task of developing a product which was demonstrated at the end of each iteration in a "sprint review". They were free to choose technology, but advised to use technology they knew. Source code was shared with the teaching staff in a repository. After demonstration, the teams delivered a retrospective report, where they reflected on their own development process and described changes for the next iteration. In total the course had seven deliverables from parts ii)-iv) which together were evaluated to provide a final grade for each team.

The previous course responsible suggested to work on improving feedback on deliverables as many students had expressed frustration on this the previous year. In a qualitative survey from 2020, students expressed that they learned a lot from the course and some stated that they "*appreciated getting feedback during the course*". However, many statements were related to the perceived quality of the feedback. One student stated that "*Uneven feedback. The teaching assistants use evaluation criteria in different ways. I asked for clarification as much which was written in the feedback was wrong, but was told that we could only complain on the whole grade and not get any more feedback*". Another commented that "*It seems like the ones giving feedback look for particular aspects not mentioned in the assignment, for example in the retrospectives. Here those who had structured the deliverable into "works well" and "could be improved" got a good score, while those who had structured the report after topics got a lower score as this was interpreted as less structured*". A third student commented that "*The feedback often seemed arbitrary; you were criticized for things not stated in the assignment such as lack of figure text. Some of the feedback is weighted very strange and seems unsupported*".

To cope with the large number of students, the teams were in the previous year given teaching assistants in several roles: One supervisor who could have regular meetings and support the team, one "product owner" who could meet with the team and provide prioritisation on features in the product. The feedback on deliverables were written by other teaching assistants. Two assistants wrote independent reports on 6-7 project team deliverables, which were again edited by a third teaching assistant before this feedback was given to the teams. To write feedback, the evaluators used a rubric [6]. For the deliverable "sprint review" at the end of first and second iterations, the rubric had 10 points which should be considered, and nine topics where evaluators were to give feedback on points which were positive and negative. In addition, the evaluator was asked to write between 100 and 200 words as feedback to the team. The third teaching assistant then prepared feedback for the student teams based on reports from independent evaluators. There were about 20 teaching assistants in 2020.

This process was time-consuming and resource-demanding. That the feedback took weeks before it was received, made it less useful for the teams, as they had already planned next steps in their project work and could be half way into the last iteration when receiving feedback on their first review and retrospective. One reason for the process with editors and independent evaluators was that there were no teaching assistants who had more in-depth knowledge on the topic of development process, which was the main focus of many of the deliverables. The editors were to ensure alignment in the feedback given to the teams. There were, however, many students who did not experience the feedback to be aligned, and who questioned the connection between the feedback and the learning objectives.

There seemed to be a large potential in improving both the timing and the quality of the feedback. In the following, I describe experience with *reorganising teaching in a semi-capstone software engineering course at bachelor level to provide project teams with more productive feedback. In particular, I will describe how we reorganised to enable feedback at scale, in a manner consistent with methods taught in the course.*

Now, I first provide background on project courses in software engineering and particular challenges in courses with providing feedback to students and managing scale in large courses. Then, I describe measures to improve feedback in 2021, and critically evaluate these measures before providing actionable advice and description of further work.

In addition to the actionable advice, this article contributes to the literature on software engineering education in showing an organisation of a large-scale software engineering project course.

## II. PROJECT COURSES, FEEDBACK AND SCALE

Many software engineering programs have elements of project work, often in the form of capstone courses [7]. Schneider et al. [8] review literature on capstone courses with emphasis on development methods introduced. Some courses train students in one method such as Scrum as in our course, others allow students to freely choose an (agile) development method. Capstone courses at master level often provide more insight in requirements engineering through a real customer, and some universities mix students at master- and bachelor-level which provides additional learning on collaboration and training. Others again have special training for Scrum masters [9].

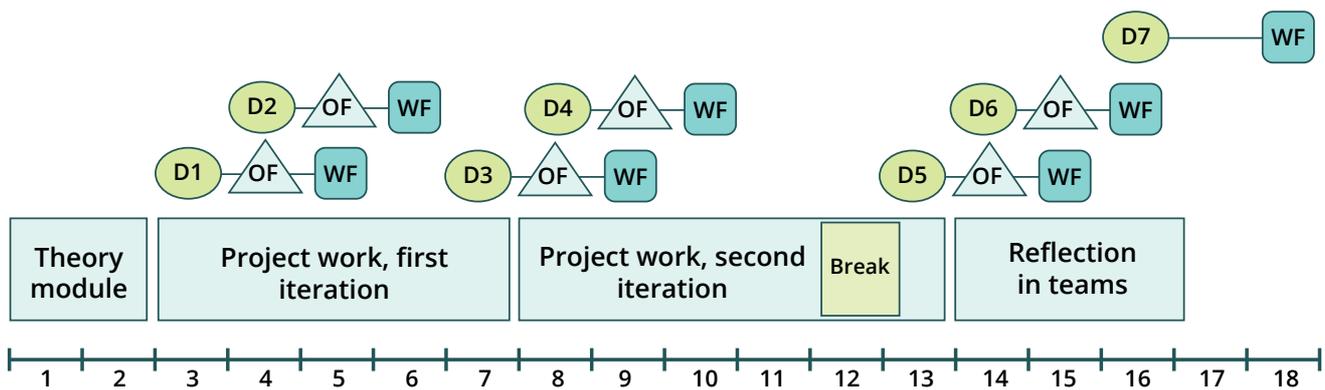

Fig. 1. Course organization with four main parts, seven deliverables (D1 – D7) with oral feedback (OF) and written feedback in analytic and holistic rubrics (WF).

An example project course is the semi-capstone course aimed to provide an "authentic learning experience" [10] which was student-centred and "delivered via active pedagogies", and "situated in a meaningful context" at Purdue Polytechnic Institute.

As with our course, iterations were used to provide student teams with formative feedback. This was also combined with traditional assessment methods.

Feedback serves a number of purposes in teaching. Price et al. [11] critically discuss feedback practice and identify five roles of feedback: Providing correction of student misconceptions, provide a positive or negative reinforcement of behaviour, forensic diagnosis, benchmarking and longitudinal development. The role that feedback has will also be influenced by when the feedback is given, how it is given, and by who. The empirical study on feedback [11] found that students in general were critical to feedback received, which was perceived as having an "overly negative tone" or was described as "vague" or "ambiguous". Teachers recognised the importance of feedback but had few ideas of the effect of the feedback apart from seeming to think that a large volume of feedback lead to increased learning.

There are also a number of approaches to ensure feedback. Esterhazy et al. [12] describes two cases of course design aimed at providing "productive feedback opportunities", with one case from biology and one from software engineering. They argue that feedback should be considered integral in course design, which should consider practices within the discipline in the design. Productive feedback opportunities is defined as giving students "opportunity to 1) make meaning of the information about the quality of their performance in the immediate course context and 2) develop skills that will allow them to capitalize on similar learning opportunities in the future". Agile software development methods such as Scrum provide several mechanisms for feedback, for reflecting in-action and on-action [13, 14]. Release planning, Sprint planning and Sprint review provide opportunities for reflection-in-action, while Sprint retrospectives and daily meetings provide opportunities for reflection-on-action.

In the pedagogical literature, many recommend the use of rubrics as a mechanism to improve feedback [6]. Rubrics can help achieving consistency in marking, provide clarity and transparency for students [15]. However, "their complexity of language may confuse students and some students do not even read rubrics because they find them too complicated" [15]. Questions have been raised in the pedagogical literature on the reliability, replicability and fairness. There has also been concern about potential disconnection between rubrics and overall course learning objectives. Bacchus et al. [15] cite a study showing that while 73% of students always read rubrics, only 43% understood what was required from reading them. However, we find many examples of successful use of rubrics, including use in project courses where one study shows a higher number of students passing a course and also a very good perception of courses amongst students [16].

But how do software engineering capstone courses manage scale? Erdogmus and Péraire describe experience from a flipped classroom software engineering course where scale was managed by increasing the number of teaching assistants, carefully recruiting teaching assistants, training of assistants and fostering a culture of mentoring amongst teaching staff [17]. In their course, a teaching assistant oversees the work of two to three teams, working ten hours per week.

Schneider et al. [8] describe the "agile capstone in education" model inspired by Spotify [18], where project teams ("squads" of 10 to 12 students, both junior and seniors) are divided into "tribes" working on a common product. At Deakin university, a teaching period had 29 project supervisors (mostly academic staff members) who assisted 302 students, organised into a variety of tribes. Supervisors have weekly supervision meetings and the model includes a three week unit with introduction for junior students and planning for senior students, followed by three iterations of product development before a project demo. The set-up is supported by a technical help-hub, and students have access to a physical collaboration work-space. As for product, an industrial representative acts as a product owner, supported by a tribe leader who translates the vision of the product owner into "manageable work packages" [8]. Feedback is given through discussions with supervisors in meetings and after each iteration, and there is also one-to-one meetings between supervisors and students to discuss progress during one week of the semester. Students compile a learning portfolio which summarises activities from the teaching period and they reflect on key learnings on how they achieve the course learning outcomes.

The model is described as providing industry-relevant experience, an authentic learning experience with effective support of students in a scalable approach [8].

### III. MEASURES TO IMPROVE FEEDBACK

Our course set-up is influenced by a long history of project courses at our department, and seeks to combine teaching of new material with practical work and reflection in a semi-capstone course design. We have chosen Scrum as main development method, as this method puts emphasis on upfront planning, involvement of the whole team in tasks, and that there are many opportunities for reflection in- and on-action as described in the previous section.

The project work assignment is given as an initial list of user stories and the product is then developed during 10 weeks with meetings which in 2020 were with one product owner and one supervisor.

We have had the following seven deliverables in the course, where the last six counted with the percentage indicated for the final grade given on the portfolio to all members who contributed on a team:

D1. Team contract (pass / not pass)

D2. Prestudy report (10%)

D3. Sprint review1 (10%)

D4. Sprint retrospective1 (10%)

D5. Sprint review2 (15%)

D6. Sprint retospective2 (15%)

D7. Team reflection report (40%)

For 2021, I suggested the following changes which were discussed with the previous course responsible and with five teaching assistants who helped plan the course:

i) Revision of assignments and rubrics in order to reduce the number of points and topics that evaluators had to consider when giving feedback.

ii) Reorganising the teaching staff in line with principles taught in the course, which shifted focus towards providing oral feedback and on increasing the speed of feedback.

iii) Extended training of teaching assistance to provide a better background for giving feedback.

I will elaborate on these three changes in the following. Figure 1 gives an overview of how the course was given in 2021 and also shows when oral and written feedback was given.

The 72 teams were formed in week 2 and had first meeting with their supervisor in week 3. They had five meetings lasting up to 45 minutes with their supervisor in the first iteration, and another five meetings in their second iteration. In all, the teaching staff conducted over 500 meetings with student teams, 144 demonstrations ("sprint reviews") and gave feedback on 504 deliverables.

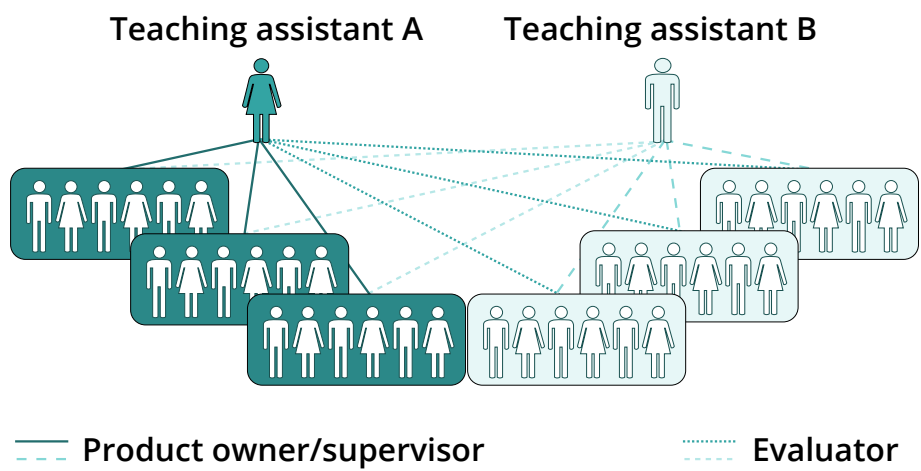

Fig. 2: Teaching assistants works in pairs within a village, each having the role of product owner and supervisor for three teams with weekly meetings, and then evaluator for deliverables from three other teams developing the same product.

In contrast to the course at Deakin university, our staff is mainly composed of teaching assistants (29), but with two lecturers and with support from four agile coaches from the software industry.

*A. Revising assignments and rubrics*

For the rubric used when teams demonstrated their product ("sprint review 1" and "sprint review 2"), the 2020 edition included many aspects such as "knowledge about the importance of a requirement specification", "understanding of the importance of communication with customer and end-user", and as mentioned 10 points to consider and nine topics where the evaluator was asked to provide feedback on strong and weak points in the presented work. It was challenging to align feedback on all these aspects.

Two teaching assistants helped to:

i) reduce the volume of text in assignments

ii) reduce the number of points and topics in rubrics in order to make it easier for evaluators to do evaluation and for students to understand feedback

iii) increase the alignment between assignment, evaluation criteria and rubric.

This was changed into a rubric with three main topics: Ability to show status of product, presentation of "user stories", and "total impression", which were stated as evaluation criteria in the assignment. The rubric included fields where the evaluator should indicate a "low", "medium" and "high" achievement on sub-topics. The "show status of product" was evaluated giving feedback on "ability to present iteration goal" and "ability to show product status through release plan, deviations from plan and consequences for further work".

*B. Reorganising the "back-end" of the course*

In 2020, the teaching assistants had the following roles:

i) supervisor for 6-7 teams

ii) product owner for 6-7 teams

iii) evaluator on deliverables for 6-7 teams

iv) editor for feedback to 6-7 teams on some deliverables

The assistants who participated in planning for 2021 said that the workload was high, and that follow-up of student teams varied with some having weekly meetings and others having meeting "at need". That a team had to relate to both a supervisor and a product owner as two persons made it challenging to have regular meetings. For all teams, the students had three other courses in parallel, and these might be different courses as students were from eight study programs.

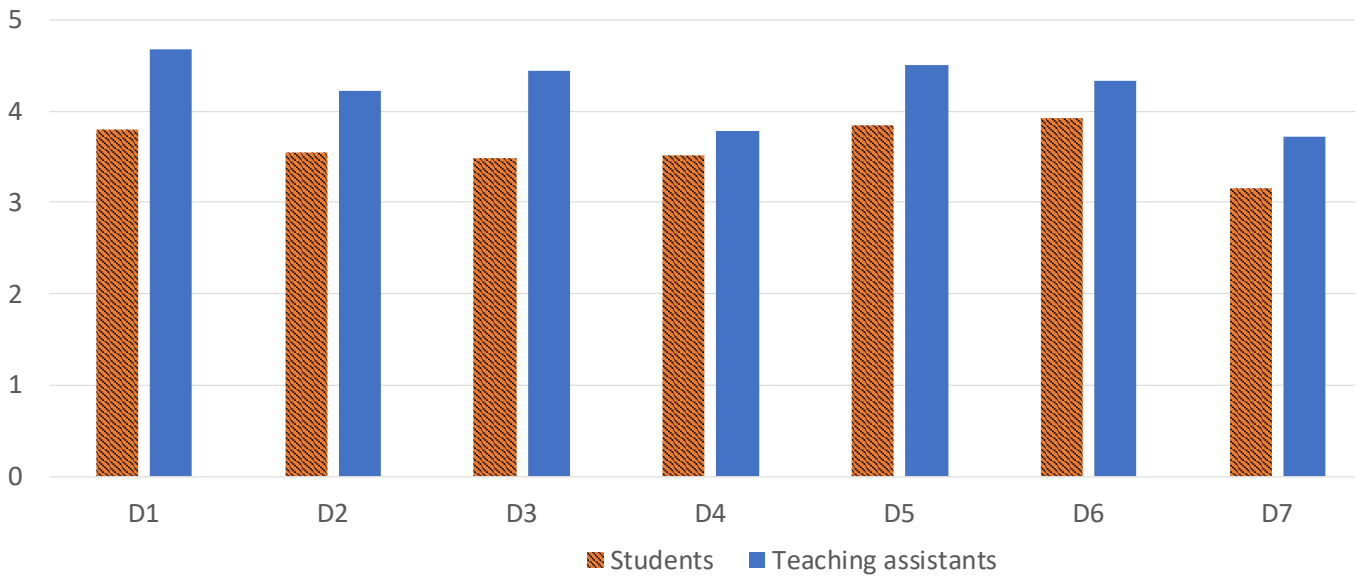

Fig. 3. Answers on question "The evaluation criteria have been clear on…" for all seven deliverables (D1-D7) in the course, answers given on a likert scale (1=totally disagree, 5=totally agree). N=142, response rate of 29% for students (orange) and N=18, response rate of 75% for teaching assistants (blue).

The set-up with independent evaluators and editors aimed to ensure alignment of feedback, but had the disadvantage that the evaluators rarely knew the student teams they were evaluating, and the supervisor was not involved in the evaluation process of own teams and might not be able to explain the feedback given to a team. The heavy focus on written and extensive feedback was not in line with principles of agile software development taught in the course, with focus on fast and oral communication [19].

To manage the challenge of scale in the course, I suggested to use inspiration from the software industry and how they manage scale in large programmes with many development teams [20]. Also, I was inspired by work in other large-scale courses at the university which managed scale in a similar manner [21]. A key principle in agile development is to rely on face-to-face communication within small teams, and the recommendation is to have a team size of 5-9 people. I suggested to use this principle also on teaching assistants so that we divided assistants into four "villages" which consisted of six assistants (three pairs as in Figure 2) and one more senior assistant as village facilitator. A village was then responsible for their own product, which meant that we did not have to align all assistants regarding knowledge of the product domain. By recruiting more teaching assistants, we were able to let one assistant supervise three teams. The supervisor would also be product owner, which meant that all communication with teams could happen in one weekly meeting. A supervisor worked together with another supervisor in the same village who acted as evaluator for his/her teams as shown in Figure 2. This meant that the evaluator did not directly meet the teams, but would get to know the team through all deliverables in the course, and also had experience with supervising three teams who developed the same product. We asked villages to have a short meeting before and after the supervision/product owner meetings had been conducted, in order to be more aligned. Supervisors were asked to mainly use time during the weekly meetings for interacting with the student team, and limit follow-up on other channels such as chat and email.

*C. Increasing training of teaching assistants*

Teaching assistants had all taken the course in previous years, and many had prior experience working in this role. Some also had experience in software development from summer jobs. All had completed a general 20-hour course on the role of being a teaching assistant.

We provided additional specific training for our course. A teaching assistant would work 100 hours, where 25 were used for training. This included five hours to recap on the course curriculum and five workshops where we discussed evaluation criteria and sometimes worked on examples from the previous year in order to align feedback. In the workshops during the semester, there were frequent discussions on "village-level" and the facilitator was also asked to conduct two retrospectives within the village in order to capture improvement suggestions on our work process in the course.

To increase motivation and insight into roles as supervisor and product owner amongst teaching assistants, we asked four professional agile coaches to join discussions at village-level in three of the workshops. We further asked one coach to present tips on working in the role as a product owner, and tips on the role of a supervisor.

## IV. EVALUATION AND DISCUSSION

Have we been successful in making changes to the course and, as Mathiassen and Puaro state, make arenas and feedback available for students to "*engage in reflections and dialogues to generate the necessary insight into the situation at hand*" [3]? Have we been able to give student teams opportunities for productive feedback [12] – that they make meaning of information about the quality of their performance and that they develop skills allowing them to capitalize on similar learning opportunities in the future?

As there were many changes in the course and as we did not physically meet students during the semester due to the pandemic, I collected feedback from students through a number of means. First, I had three meetings with a reference group consisting of 12 students who volunteered to meet and discuss the course. In collaboration with the reference group, I conducted two surveys to all students, one after the second iteration of project work and one after the course had finished but before teams were given a final grade. The teaching assistants participated in two retrospectives in each village during the semester, completed a questionnaire after the work was finished and also wrote an experience report from each village.

In general, the course was perceived to provide a good learning outcome. In the surveys, we asked students to indicate agreement on a likert scale (1=totally disagree, 5=totally agree). On the second survey to students (N=142), the score for "I have learned much about software engineering" was 3.9 and on "in total, how satisfied are you with the course" got 3.6. The question "the course content will be relevant for work" got 4.2.

The final grades given shows a high achievement of learning objectives in that about 30% of the students got an A, 65% a B and 5% a C.

We asked how much time students were working on each part of the course, and they self-report to use much more than expected. We expect 12 hours of effort per week, while students reported on average 8.3 on theory module, 11.5 on iteration 1, 13.7 on iteration 2 and 15.7 on reflection in teams.

Also, we measured the team cohesion through three questions and the score was 4.1 which indicates that we were successful in providing a good structure for project work despite of all parts of the course being online.

Comparing the course in 2021 to findings in the literature, we first note that project work seems to give much motivation to students. The project work seems to bring positive effects from giving effort to work in the course to also give competence perceived as relevant for work-life.

### A. Perception of feedback

To understand more about how feedback was perceived, I asked the question "did you receive useful feedback" for each of the first six deliverables where teams had gotten feedback at the time of the survey. The scores are in the range from 4.1 to 4.4 with feedback on "retrospective 2" getting the highest score. I also asked if the evaluation criteria were clear, and for this question the scores varied from 3.2 to 3.9, again with "retrospective 2" getting the highest score (see Figure 3). As shown in the Figure 3, there is an increase in perceived clarity of the criteria on the deliverables which were repeated, the "sprint review 1 and 2" and "retrospective 1 and 2". The last deliverable involved a new task we had not trained on, which was to conduct an own literature search and identify relevant articles for a selected sub-topic, which could explain the relatively low score (3.2).

The quantitative findings on quality of feedback indicate that the quality is reasonably high. However, looking into qualitative statements we find some positive comments on feedback such as "*we have received good feedback in the course*", and "*it was good to have more of the same type of deliverables, so that we can improve*" and "*useful and fast feedback on the deliverables*". But we also find a number of comments similar to what was given the previous year such as improvement suggestions on "*make evaluation criteria clearer – we got a low score on something we did not know was evaluated*" and "*some teaching assistants seemed to be much more strict than others, which makes the assessment unfair*", and "*the teaching assistant give strange feedback and does not want to discuss the feedback. The feedback depends a lot on which teaching assistant you get*".

By reorganising the course, we were able to give teams more oral feedback, and a weekly opportunity to receive feedback and advice on ongoing work. The oral feedback was given by the supervisor, who knew the team, their context and product. In line with findings [11], we have tried to shorten the time between a deliverable and when feedback was given by first giving an oral feedback on deliverables one to six. As shown in Figure 1, the feedback was given first orally after a week and then in written two weeks after delivery of an assignment. One could argue that two weeks for written feedback is still too long, as teams are quite far into the next iteration when receiving feedback.

We tried to avoid vague feedback by using analytic and holistic rubrics, and also tried to ensure that student teams understood the feedback as it was first explained and then later provided in written. As Esterhazy et al. [12] describe, we exploited opportunities within the discipline in making a course design giving feedback aligned with practices in the software industry. With the current design, we depend on the proper training of the teaching assistants, but in most cases seem to achieve that teams make meaning of information about the quality of their performance, and also that they are able to develop skills for future similar learning opportunities for "sprint reviews" and "retrospectives".

In the pedagogical literature, many studies focus on the use of rubrics as a mechanism to improve feedback, such as [16] where a combination of analytic and holistic rubrics are used as in our course. With our efforts to revise rubrics and the feedback process, the survey results indicate that it is not a large problem for student teams to understand the feedback as found in previous studies [15]. However, students raise the question of fairness, both in that they perceive some teams to get more feedback from their supervisor, and that some teams get a more strict evaluator. Given the resources in the course, we will probably not be able to provide more training to assistants, but we can organise workshops to focus more on alignment of feedback. Further, we could also try to recruit assistants who have better background skills and possibly also better motivation to act as a supervisor.

*B. Managing scale*

The main difference with this course compared to my previous teaching experience has been the scale. I have taught a smaller project course using approximately the same set-up with four main parts of the course and focus on formative assessment through early deliverables [22]. In that course, we also developed rubrics in collaboration with all teaching assistants. In the current course, this work was done with involvement from five assistants who participated in planning and not all assistants. Involving teaching assistants could probably increase understanding of evaluation criteria and quality of feedback further, but as reported in the survey this is perceived to be of a reasonably high quality. This is also an activity which would take time, and given the challenge with scale it might be difficult to develop a rubric with involvement of the high number of assistants.

Teaching assistants give very good feedback on the organisation of work into "villages" with a score of 4.9. An assistant stated: "*very satisfied with the organisation of the staff. It has been really easy to ask questions if I have been uncertain, and I have never been uncertain about what is to be done at any time. Very impressed with all information and organisation*". Another stated "*the collaboration with the partner has worked well on evaluations. You got other input on things you did not see yourself*".

The feedback that assessment was "unfair" as student teams perceived they got different levels of support from assistants and that assistants were more of less strict was also a critique in the previous project course. It would be interesting to investigate further what might be reasons for the perception of unfairness, and if we could reduce this perception through more information about what can be expected from the teaching staff or from increasing training of teaching assistants to make assistants comfortable giving feedback to teams. We also have unused opportunities in involving students in evaluating other teams, this is something we have not had resources to plan yet, but are considering for 2022. To reduce the impact of a teaching assistant, the assessment will from 2022 include an individual component which is evaluated by others than the teaching assistants and is part of a summative assessment.

*C. Perception of training*

The teaching assistants in general give higher scores on the clarity of evaluation criteria (3.4 to 4.7 – see Figure 3) and also on average answer that they have received good support from the staff in their work as evaluators (4.3 – 4.5). One assistant stated that "*the workshops have given me as a teaching assistant a foundation to help the teams in their process, and they have helped when evaluating*". Overall, the teaching assistants gave a score of 4.6 to the statement "training on evaluating has been useful". The score for "I have received support in my role as supervisor" got 4.3 and "I have received support in my role as product owner" 4.0. An assistant commented that "*you learn and develop as teaching assistant, which means that you gain something beyond a salary. This is very good and not the case of most teaching assistant positions*". Most teaching assistants would recommend the job to someone else (4.3).

V. ACTIONABLE ADVICE AND FUTURE WORK

I have presented the challenge of providing high-qualify feedback in a bachelor-level project course in software engineering with about 500 students, organised into 72 teams. The course is generally well-received by students, and students invest much time. I have discussed the course design and changes in 2021 to increase quality and provide productive feedback, which included revising assignments and rubrics, reorganising the teaching staff into "villages" responsible for a product domain, and increasing training of teaching assistants. Based on feedback from students, teaching assistants, findings in literature and own experience, I have discussed how the three changes impacted on the quality of feedback.

I propose the following advice for other large-scale software engineering courses seeking to improve feedback to teams in project work:

- *Sacrifice breadth and volume in feedback to increase timeliness and fairness:* Revising assignments and rubrics to cover only most important themes, and first providing oral and then written feedback led to timely and high-quality feedback which was perceived useful by the student teams. Our set-up exploits opportunities in agile development for in-action and on-action reflection as suggested in [12]. Feedback on other themes is given orally in weekly supervision meetings.

- *Exploit lessons learned in industry on how to manage scale:* Organising teaching assistants in villages with focus on one product reduces the need for knowledge-sharing across the whole course staff and empowers teaching assistants in the role of product owner.

- *Invest in training of teaching assistants to improve the learning outcome of project work:* Supervising a development team and acting as a product owner are challenging tasks ideally done by people with development experience and domain knowledge. Training should focus both on the core topics covered in the course as well as on how to evaluate deliverables and give feedback.

The quality is generally good, but there is still a perception amongst students that assessment is "unfair" and that it is dependent on the teaching assistant. For 2022, we will seek to overcome this challenge by better informing about what could be expected from teaching staff, continue to train teaching assistants, focus more on background knowledge and suitability in supervision role when recruiting, and finally reduce the importance of potential differences in evaluation in project work by including a new individual component in the assessment.

ACKNOWLEDGMENT

I would like to thank Dagrun Astrid Aarø Eggen and Letizia Jaccheri at the Norwegian University of Science and Technology for comments on earlier versions of this article, as well as six teaching assistants who also provided comments: Anniken Grimsmo, Tarjei Johre, Mattias Ness, Sindre Langaard, Peder Smith and Catherine Xu.

REFERENCES

[1] Nytrø, Ø., Nguyen-Duc, A., Trætteberg, H., Lorås, M., and Farschian, B. A., "Unreined Students or Not: Modes of Freedom in a Project-Based Software Engineering Course," in *2020 IEEE 32nd Conference on Software Engineering Education and Training* (CSEE&T), 2020, pp. 1-10.
[2] Kolås, L. and Munkvold, R. I., "Learning through construction: a roller coaster ride of academic emotions?," in P*roceedings of the 6th Computer Science Education Research Conference*, 2017, pp. 10-19.
[3] Mathiassen, L. and Purao, S., "Educating reflective systems developers," *Information Systems Journal,* vol. 12, pp. 81-102, 2002.
[4] Kniberg, H., Scrum and XP from the Trenches, 2nd edition ed.: InfoQ, 2015.
[5] Sommerville, I., Software Engineering, Tenth edition ed. Harlow, England: Pearson Education Limited, 2016.
[6] Ambrose, S. A., Bridges, M. W., DiPietro, M., Lovett, M. C., and Norman, M. K., How learning works: Seven research-based principles for smart teaching: John Wiley & Sons, 2010.
[7] Cico, O., Jaccheri, L., Nguyen-Duc, A., and Zhang, H., "Exploring the intersection between software industry and Software Engineering education - A systematic mapping of Software Engineering Trends," *Journal of Systems and Software,* vol. 172, p. 110736, 2021/02/01/ 2021. https://doi.org/10.1016/j.jss.2020.110736
[8] Schneider, J.-G., Eklund, P. W., Lee, K., Chen, F., Cain, A., and Abdelrazek, M., "Adopting industry agile practices in large-scale capstone education," in *Proceedings of the ACM/IEEE 42nd International Conference on Software Engineering: Software Engineering Education and Training,* 2020, pp. 119-129.
[9] Paasivaara, M., "Teaching the Scrum Master Role using Professional Agile Coaches and Communities of Practice," in *2021 IEEE/ACM 43rd International Conference on Software Engineering: Software Engineering Education and Training* (ICSE-SEET), 2021, pp. 30-39.
[10] Magana, A. J., Seah, Y. Y., and Thomas, P., "Fostering cooperative learning with Scrum in a semi-capstone systems analysis and design course," *Journal of Information Systems Education,* vol. 29, p. 4, 2019.
[11] Price, M., Handley, K., Millar, J., and O'donovan, B., "Feedback: all that effort, but what is the effect?," *Assessment & Evaluation in Higher Education,* vol. 35, pp. 277-289, 2010.
[12] Esterhazy, R., Nerland, M., and Damşa, C., "Designing for productive feedback: an analysis of two undergraduate courses in biology and engineering," *Teaching in Higher Education*, pp. 1-17, 2019.
[13] Babb, J., Hoda, R., and Norbjerg, J., "Embedding reflection and learning into agile software development," *IEEE Software*, vol. 31, pp. 51-57, 2014.
[14] Burden, H. and Steghöfer, J.-P., "Teaching and Fostering Reflection in Software Engineering Project Courses," in A*gile and Lean Concepts for Teaching and Learning*, ed: Springer, 2019, pp. 231-262.

[15] Bacchus, R., Colvin, E., Knight, E. B., and Ritter, L., "When rubrics aren't enough: Exploring exemplars and student rubric co-construction," *Journal of Curriculum and Pedagogy,* vol. 17, pp. 48-61, 2020.

[16] Martínez, F., Herrero, L. C., and De Pablo, S., "Project-based learning and rubrics in the teaching of power supplies and photovoltaic electricity," *IEEE Transactions on Education,* vol. 54, pp. 87-96, 2010.

[17] Erdogmus, H. and Péraire, C., "Flipping a graduate-level software engineering foundations course," in *2017 IEEE/ACM 39th International Conference on Software Engineering: Software Engineering Education and Training Track* (ICSE-SEET), 2017, pp. 23-32.

[18] Edison, H., Wang, X., and K., C., "Comparing Methods for Large-Scale Agile Software Development: A Systematic Literature Review," *IEEE Transactions on Software Engineering*, pp. 1-1, 2021. 10.1109/TSE.2021.3069039

[19] Baham, C. and Hirschheim, R., "Issues, challenges, and a proposed theoretical core of agile software development research," *Information Systems Journal*, vol. n/a, 2021. https://doi.org/10.1111/isj.12336

[20] Dingsøyr, T., Falessi, D., and Power, K., "Agile Development at Scale: The Next Frontier," *IEEE Software,* vol. 36, pp. 30-38, 2019. 10.1109/MS.2018.2884884

[21] Veine, S., Anderson, M. K., Andersen, N. H., Espenes, T. C., Søyland, T. B., Wallin, P., and Reams, J., "Reflection as a core student learning activity in higher education - Insights from nearly two decades of academic development," *International Journal for Academic Development,* vol. 25, pp. 147-161, 2020/04/02 2020. 10.1080/1360144X.2019.1659797

[22] Dingsøyr, T., "Agile Iteration Reviews in a Project Course: A key to Improved Feedback and Assessment Practice," in *Workshop on Software Engineering Education for the Next Generation, International Conference on Software Engineering*, Madrid, Spain (online), 2021, pp. 21-25.